\documentstyle[12pt]{article}
\newlength{\bredde}
\def\slash#1{\settowidth{\bredde}{$#1$}\ifmmode\,\raisebox{.15ex}{/}
\hspace*{-\bredde} #1\else$\,\raisebox{.15ex}{/}\hspace*{-\bredde} #1$\fi}
\newcommand{\beq}{\begin{equation}}
\newcommand{\eeq}{\end{equation}}
\newcommand{\bea}{\begin{eqnarray}}
\newcommand{\eea}{\end{eqnarray}}

\def\gtwid{\raise.3ex\hbox{$>$\kern-.75em\lower1ex\hbox{$\sim$}}}
\def\ltwid{\raise.3ex\hbox{$<$\kern-.75em\lower1ex\hbox{$\sim$}}}
\begin{document}
\topmargin -0.8cm
\oddsidemargin -0.8cm
\evensidemargin -0.8cm
\title{\Large{Smooth Non-Abelian Bosonization}}

\vspace{0.5cm}

\author{{\sc P.H. Damgaard} \\
CERN -- Geneva \\~~\\and \\~~\\
{\sc R. Sollacher} \\
Gesellschaft f\"{u}r Schwerionenforschung GSI mbH \\
P.O. Box 110552, D-64220 Darmstadt, Germany
}
\maketitle
\vfill
\begin{abstract}
We present an extension of ``smooth bosonization'' to the non-Abelian
case. We construct an enlarged theory containing both bosonic and
fermionic fields which exhibits a local chiral gauge symmetry. A gauge
fixing function depending on one real parameter allows us to
interpolate smoothly between a purely fermionic and a
purely bosonic representation. The procedure is, in the
special case of bosonization, complementary to the approach based
on duality.
\end{abstract}
\vfill
\vspace{6.5cm}
\begin{flushleft}
CERN--TH-7347/94 \\ hep-th/9407022 \\
July 1994
\end{flushleft}
\newpage

\setcounter{page}{1}

\section{Introduction}

Whereas $(1\!+\!1)$-dimensional bosonization of one fermion species,
Abelian bosonization, has a
long history (in field theory language most compactly expressed in the
form given in ref. \cite{Abelian}), a proper bosonization of fermionic
theories with
internal symmetries  --  or just more species -- was only achieved ten years
ago \cite{Witten}. The trouble is that if one naively applies the Abelian
bosonization rules, one obtains a form of the action that is not
manisfestly symmetric under the same non-Abelian transformations as the
fermionic theory. Witten solved this problem in a very elegant
way by showing that a free theory of $N$ Dirac fermions is equivalent
to a bosonic $O(2N)\!\times\!O(2N)$-symmetric sigma model with a Wess-Zumino
term \cite{Witten}. A precursor of this equivalence can be found in the
work of Polyakov and Wiegmann \cite{Polyakov}, and path-integral derivations
of the non-Abelian fermi-bose equivalence in two dimensions have been
given in ref. \cite{DiVecchia}. For a nice review, see also the textbook
\cite{Abdalla}.

Surprisingly, Abelian bosonization can be been shown to be but a
very special case within a huge class of equivalence \cite{us}. This
generalization of the conventional Abelian bosonization prescription is
achieved by first finding a gauge-symmetric theory (with a particular chiral
gauge symmetry that will be further explained below) containing both bosons
and fermions. Two gauge fixings of this ``higher'' gauge-symmetric theory
turn out to correspond to a description entirely in terms of fermions
or entirely in terms of bosons. Bosonization (or
fermionization) is thus nothing but a question of choosing a specific
gauge in a particular gauge-symmetric theory of both bosons and fermions. More
importantly, one can find a smooth gauge, parametrized in terms of one real
variable $\Delta$ such that one moves smoothly from a fermion theory (reached
at $\Delta = 0$) to a boson theory (reached at $\Delta = 1$) \cite{us}.
At all values
of $\Delta$ in between, one has neither a purely fermionic nor purely bosonic
description, but a mixed representation of apparently interacting bosons
and fermions. All of these theories are equivalent.  We have called this
phenomenon {\em smooth bosonization}.

Very recently, Burgess and Quevedo \cite{Burgess} have shown
that a natural generalization of duality transformations (as known in
string theory or two-dimensional conformal field theory) in
$(1\!+\!1)$-dimensional fermionic or bosonic theories can map one theory into
the other. Their proposal for the appropriate generalization of the notion
of duality is to take the route of Ro\v{c}ek and Verlinde \cite{Rocek}, and
gauge a vector-like symmetry of the fermionic theory. After introducing
a Lagrange multiplier to constrain one of the new degrees of freedom
by fixing on flat connections, a genuine gauge can be chosen for the remaining
single gauge degree of freedom. Upon integrating out the fermions and the
gauge potential of the functional integral, the resulting theory of
the Lagrange multiplier is precisely the appropriate bosonized theory.
The bosonic and fermionic theories are thus, in this precise technical sense,
really dual to each other. This approach is interesting, because it yields
a complementary means of applying the same idea as in ref.
\cite{us}. It does not seem, however, to lead easily to the kind of
generalizations described in ref. \cite{us}. In detail, the
difference between the approach known as smooth bosonization and that based
on duality can be described as follows. The duality transformation changes
variables inside the functional integral by going from a fermionic field to a
Lagrange multiplier, which ends up as the surviving bosonic field. The gauge
degrees of freedom, as well as the original fermionic degrees of freedom, are
explicitly integrated out of the functional integral. In contrast, smooth
bosonization hinges on a change of variables in the
functional integral, where the fermionic degrees of freedom are never
integrated out of the path integral. Instead, they decouple in the ``boson
gauge'' from the relevant Green functions. These ghostly remains of the
fermions in the bosonic formulation are not entirely vanishing.
They are fermions without currents (neither vector nor axial vector currents,
one implying the other due to the two-dimensional identity $\gamma_\mu
\gamma_5 \!=\! - \epsilon_{\mu\nu}\gamma^\nu$, and
hence have no charges. In any other gauge except this bosonic gauge they
still carry non-trivial degrees of freedom. The bosonic degree of freedom
in smooth bosonization is not the Lagrange multiplier of a gauge constraint,
but rather the single gauge degree of freedom itself. Even in the particular
case of the boson gauge ($\Delta \!=\! 1$), smooth bosonization is thus
complementary to, and not identical to, the approach to bosonization based
on duality.

Also non-Abelian bosonization can be
given an interpretation in terms of duality transformations (see the
second paper of ref. \cite{Burgess}). Here the notion of duality has to
be somewhat enlarged to a non-Abelian context \cite{Ossa}, but the
procedure is otherwise the same. Again, the duality transformation does not
appear to provide any new information on the fermi-bose equivalence,
but it reproduces all known results. It is also the first time a step-by-step
derivation of the appropriate non-Abelian bosonized theory, obtained
directly by path-integral manipulations of the fermionic theory.

In view of these recent developments, the odds would seem to be in favour
of a possible generalization of smooth bosonization to include the
non-Abelian case. What we are seeking, then, is a gauge-symmetric theory
of both fermions and bosons (with appropriate indices to make the action
invariant also under global non-Abelian rotations). This theory should be
constructed in such a way that a gauge exists which smoothly takes us from a
fermionic theory to a bosonic theory as we vary a single parameter $\Delta$.
This would give the sought-for generalization of non-Abelian
bosonization. The aim of this paper is to provide an appropriate
gauge-symmetric theory with these properties, and a smooth gauge which
interpolates between fermionic and bosonic descriptions.

Such a smooth non-Abelian bosonization is of interest also from another
point of view. Since both Abelian and non-Abelian bosonization can be
thought of as obtained by applying duality transformations, the existence
of a generalization of this notion to a {\em continuous} class of
transformations whose endpoints coincide with the usual duality
transformations (here fermi-bose transmutations) would strongly suggest that
also the concept of duality in conformal field theory is open to
generalization. As we will demonstrate here, it is indeed possible to
generalize also non-Abelian bosonization to a smooth, larger, version.
There thus seems to be increasing evidence that
conventional duality transformations, be they Abelian or non-Abelian,
may form only the endpoints of a
much larger class of mappings that bring one theory into an equivalent
one.\footnote{See also the discussion in section 2 of ref. \cite{Alvarez}.}

Finally, obtaining the correct prescription for smooth non-Abelian
bosonization in $(1\!+\!1)$-dimensions is a much needed ingredient in defining
suitable semi-bosonized versions of low-energy effective Lagrangians
in $(3\!+\!1)$ dimensions. Just as one can derive an effective long-distance
action for the $\eta'$ degree of freedom in QCD \cite{us1}, one can apply
the techniques
of smooth non-Abelian bosonization to derive an effective quark-meson
Lagrangian for the $SU(N_f)$ pseudoscalar multiplets of $N_f$-flavour QCD.
But such a phenomenological application of our formalism, partial bosonization
of regularized QCD, obviously lies much beyond the scope of the present paper.

\section{Finding the gauge-symmetric theory}

The first step in establishing the existence of smooth non-Abelian
bosonization consists in finding an appropriate gauge-symmetric theory
of both bosons and fermions. Guided by the experience gained in the
abelian case \cite{us}, we will derive this gauge-invariant theory by
a collective field technique based on a local non-Abelian chiral
rotation of fermions \cite{Alfaro}.

Our conventions are as follows.
We consider $N$ species of Dirac fermions in Minkowski space:
\beq
\psi = (\psi_1,\ldots ,\psi_N) ~.
\eeq
The functional integral governing the dynamics of these fermions is
chosen to be
\bea
Z[V,A]_\Lambda &=& \int\! {\cal D}[\psi,\bar{\psi}]_\Lambda \; e^{iS_\psi} \cr
{\cal S}_\psi &=& \int\! d^2 x\; \bar{\psi} i \slash{D} \psi \cr
D_\mu &=& \partial_\mu -iV_\mu -iA_\mu\gamma_5 ~.
\label{eq:Zf}
\eea
Here, $V_\mu = V_\mu^A T_A$ and $A_\mu = A_\mu^A T_A$ are
external sources; the $T_A$'s are the generators of a Lie group $SU(N)$
or $U(N)$. By adding suitable terms depending only on these sources, and
integrating over the sources in the functional integral, one can generate
a number of non-trivial theories (non-Abelian Thirring models, non-Abelian
gauge theories with fermions, etc.). So the generating functional
(\ref{eq:Zf}) is sufficiently general for our purpose.\footnote{Mass terms
lead to a far more complicated procedure, and will not be discussed in this
paper. Compare also the different level of difficulty in deriving smooth
Abelian bosonization with and without mass terms \cite{us}.}

In order to deal with a well-defined functional integral in
(\ref{eq:Zf}) we must at least impose a regularization on the fermionic path
integral, here indicated by the subscript $\Lambda$ which is the
ultraviolet cut-off. We choose a
consistent regularization scheme like the Pauli-Villars scheme
described in ref. \cite{Ball}. It explicitly preserves vector gauge
symmetry, here non-Abelian phase rotations of the fermions.
When mass terms are not present, one can
in the end send the cut-off $\Lambda$ to infinity in a straightforward
way (see below).

It is convenient to introduce projectors on definite chirality,
\beq
P_\pm = \frac{1}{2} (1 \pm \gamma_5)~~,
\eeq
as well as light-cone components for an arbitrary two-vector $a_\mu$:
\beq
a^\pm = \frac{1}{\sqrt{2}} (a^0 \pm a^1)~.
\eeq
Using the relations
\beq
\gamma^+ P_+ = \gamma^+ ~~,~~\gamma^- P_- = \gamma^- ~~,~~ \gamma^+
P_- = 0 ~~,~~ \gamma^- P_+ = 0 ~,
\eeq
the action simplifies:
\beq
{\cal S}_\psi = \int\! d^2 x\; \bar{\psi} i (\gamma^+ P_+ D^L_+ +
\gamma^- P_- D^R_- )\psi ~.
\label{eq:Spsi}
\eeq
Here we have defined
\beq
D^L_\mu = \partial_\mu -iL_\mu ~=~ \partial_\mu -iV_\mu -iA_\mu
\eeq
and
\beq
D^R_\mu = \partial_\mu -iR_\mu ~=~ \partial_\mu -iV_\mu +iA_\mu~~.
\eeq
Only two couplings to external sources are actually playing a r\^{o}le,
namely those to $L_+$ and to $R_-$. This is the light-cone analogue of
the relation
$\gamma_\mu\gamma_5 = - \epsilon_{\mu\nu}\gamma^\nu$ between
$\gamma$-matrices in two dimensions.

Having fixed our notation, we now proceed by extending the system through
the introduction of appropriate collective fields. We are of course guided
by the experience gained in the Abelian case, where the appropriate field
transformation involves a chiral rotation. In this case the natural
procedure is to extend this to a non-Abelian chiral transformation.
The bosonic fields are then matrices
$U(x)$ being elements of a group $SU(N)$ or $U(N)$, $i.e.$,
\beq
U(x) = e^{2i\theta (x)} ~~, ~~ \theta (x) = \theta^A T_A~~.
\eeq
In the Abelian case \cite{us} this extension of the field space was
achieved by introducing a pseudoscalar field via a straightforward chiral
transformation of the fermion fields. One can follow the same procedure
in this non-Abelian case, but we choose for convenience to depart slightly
from this route here, and introduce the fields $U(x)$ by a transformation
involving only one chiral component of $\psi$:
\beq
\psi (x) = (U(x) P_+ + P_-)\chi (x)~~, ~~\bar{\psi} (x) =
\bar{\chi} (x) (U^\dagger (x) P_- + P_+ )
\label{eq:trans}
\eeq
This transformation differs from a purely chiral rotation by an additional
(non-Abelian) phase transformation $\chi (x) \to \exp (i\theta (x))
\chi (x)$.

It is well known that a chiral transformation like (\ref{eq:trans})
has a non-trivial Jacobian due to
the regularized fermionic measure in the functional integral.
In ref. \cite{Ball} one can find an expression for the Jacobian associated with
this transformation (\ref{eq:trans}):
\bea
\log J &=& \int_M\! d^2x\; \biggl( \frac{1}{8\pi} tr\; \partial_\mu U
\partial^\mu U^\dagger
+ \frac{i}{4\pi} tr\; U^\dagger L_\mu U ( U^\dagger \partial^\mu U +
\epsilon^{\mu\nu} U^\dagger \partial_\nu U ) \cr
&& - \frac{i}{4\pi} tr\; R_\mu ( U^\dagger \partial^\mu U -
\epsilon^{\mu\nu} U^\dagger \partial_\nu U )
-\frac{1}{4\pi} tr\; (R_\mu + \epsilon_{\mu\nu} R^\nu ) (U^\dagger
L^\mu U - L^\mu ) \biggr)\cr
&&+ \epsilon_{\mu\nu\rho} \frac{1}{12\pi} \int_B d^3x\; tr\; U^\dagger_s
\partial^\mu U_s U^\dagger_s \partial^\nu U_s, U^\dagger_s
\partial^\rho U_s + {\cal O}(\Lambda^{-2})~~.
\eea
The first integral is over 2-dimensional space-time, the second ---
the Wess-Zumino term ---
is over a 3-dimensional manifold with space-time M as its boundary.
This expression corresponds to a local term in
the new action, proportional to $\hbar$ if we re-instate factors of $\hbar$.
In the limit where the ultraviolet cut-off is sent to infinity and in
terms of light-cone coordinates, the result is simple:
\bea
Z[V,A] &=& \int\! {\cal D}[\chi,\bar{\chi}] \;
e^{i\int\! d^2x\; {\cal L}_\chi} \cr
{\cal L}_\chi &=& \bar{\chi}i (\gamma^+ P_+ D^{L,U}_+ +
\gamma^- P_- D^R_- ) \chi  +\frac{1}{4\pi} tr\; \partial_+ U \partial_-
U^\dagger \cr &&+ \frac{1}{4\pi} \int_0^1 \! ds\; tr\; 2i\theta [U^\dagger_s
\partial_+ U_s, U^\dagger_s \partial_- U_s ]
-\frac{i}{2\pi} tr\; L_+ U\partial_- U^\dagger \cr &&-\frac{i}{2\pi} tr\;
R_- U^\dagger \partial_+ U
-\frac{1}{2\pi} tr\; R_- (U^\dagger L_+ U - L_+) ~.
\label{eq:Ztr}
\eea
Here we have used the abbreviations
\beq
U_s = e^{2is\theta}
\eeq
and
\beq
D_+^{L,U} = \partial_+ -i U^\dagger L_+ U + U^\dagger \partial_+ U~~.
\eeq
For finite cut-off $\Lambda$, there are
additional terms of order ${\cal O} (\Lambda^{-2})$ which contain higher
derivatives of $U(x)$, or higher powers of the external sources
suppressed by suitable powers of the cutoff $\Lambda$. We will argue
later that for the following one can neglect the effect of
these terms in the limit of infinite cutoff.

Now we can extend the theory, treating the collective field
$U(x)$ as a quantum field. This
means that the functional integral (\ref{eq:Ztr}) is averaged with
respect to all possible configurations $U(x)$:
\beq
Z_{ext} [V,A]  = \int\! {\cal D} [U]\; Z[V,A]
\label{eq:Zext}
\eeq
As $Z[V,A]$ is independent of $U(x)$ such a manipulation just
introduces an overall volume factor. Associated with this degeneracy
is a local gauge symmetry. We choose to integrate $U(x)$ over the Haar
measure. The partition function is then invariant under the local gauge
transformation
\bea
\chi (x) &\to & (A(x) P_+ + P_-)\chi (x)\cr
\bar{\chi} (x) &\to & \bar{\chi} (x) (A^\dagger (x) P_- + P_+ )\cr
U(x)&\to & U(x) A^\dagger (x) \cr
U^\dagger (x) &\to& A(x) U^\dagger (x)
\label{eq:sym}
\eea
Here, $A(x)$ is a unitary matrix belonging to the same Lie group as $U(x)$.
Note that neither the action nor the functional measure are separately
invariant under this symmetry, but the combination is.

\section{A smooth gauge between fermions and bosons}

We have now achieved our first goal. We have found a new way of defining our
fermionic theory so that it involves a non-trivial collection of bosonic
fields as well. This means that we are now in a position to choose between
different representations. It can be achieved by including a suitable
$\delta$-functional inside the path integral --- or in a different language,
by choosing a
gauge. We do it by imposing as many gauge constraints as there are new field
degrees of freedom. In the present case this number is equal to the
number of generators of the group, $i.e.$ $N^2$ for $U(N)$ or $N^2-1$
for $SU(N)$.

Our aim is to provide a gauge which interpolates smoothly between a fermionic
and a bosonic representation. Glancing at the transformed action in
(\ref{eq:Ztr}) one observes that this certainly can be achieved for at
least one quantity, namely one of the currents coupling to the
external sources. For example, by taking a derivative with respect to the
source $L_+$, we find, inside the path integral, the following shift due to
the introduction of the collective field $U(x)$:
\beq
\bar{\psi} \gamma^+ P_+ T_A \psi = \bar{\chi} \gamma^+ P_+ U^\dagger
T_A U \chi - \frac{i}{2 \pi} tr\; T_A U D^R_- U^\dagger ~.
\label{eq:shift}
\eeq

Although the whole physical current --- the left hand side of eq. (16) ---
is gauge invariant, each of the transformed
components in eq. (\ref{eq:shift}) are not. As in the Abelian case
\cite{us}, we now  try to choose a gauge
such that the bosonic part describes a fraction $\Delta$ of the whole
physical current. This is not entirely trivial, since the addition of the
gauge-fixing constraint in itself modifies the regulator, which in turn,
for consistency, modifies the Jacobian associated with the non-Abelian
chiral transformation. In ref. \cite{us} we called this phenomenon
``anomalous gauge fixing'', because in the Abelian case the pertinent
gauge fixing turned out to hinge directly on the $U(1)$ anomaly in
$(1\!+\!1)$ dimensions. Actually, the phenomenon is more general,
and not necessarily linked to anomalies. It occurs whenever the
gauge-fixing function causes modifications at the quantum level (here,
at the one-loop level, because it occurs directly as a consequence of
a non-trivial Jacobian, of order $\hbar$ when exponentiated into the action).
So a more apt name would be
``gauge fixing at the quantum level''. In any case, for a situation
similar to the present,
we have already provided a fairly straightforward recipe in ref. \cite{us2}.
The idea is to choose a more conventional gauge (in this case, entirely in
the bosonic $U(x)$-sector) where no modification of the regulator is
required. Since, on the other hand, this does not provide the
gauge we are seeking, one more ingredient is needed. The way
to move into the more interesting gauges from one of these trivial
or ``classical'' ones, is to shift the external sources by the same
Lagrange multiplier that enforces the gauge constraint. One can easily
show, and it will be demonstrated explicitly below on the basis of BRST
invarience, that this is a correct procedure.

Using a Lagrange multiplier field $b = b^A T_A$, we hence both shift the source
$L_+$ by the amount $\Delta b$ ($\Delta$ being an arbitrary real number,
the parameter which will turn out to smoothly join fermionic and bosonic
gauges) and add a term
$$
\frac{i}{2\pi} tr\; b U D^R_- U^\dagger~~
$$
to the action.

This is not the full story, however. We find the associated
Faddeev-Popov determinant by performing an infinitesimal gauge variation.
Expressed in terms of conventional non-Abelian ghost fields, we get
\beq
{\cal L}_{ghost} = \frac{1}{2\pi} tr\; U^\dagger \bar{c} U D_-^R c~~,
\eeq
where $\bar{c} = \bar{c}^A T_A$ and $c = c^A T_A$ are the Grassmann-odd
ghost fields (in the adjoint representation). Covariant derivatives
acting on $c$ are correspondingly defined as
\beq
D_-^R c = \partial_- c -i[R_-,c]~~.
\eeq
Similar relations hold for $U,b$ and $\bar{c}$.

With these ingredients the original functional integral now reads:
\bea
Z[V,A] &=& \int\! {\cal D} [\chi, \bar{\chi}] {\cal
D}[U] {\cal D} [b] {\cal D} [c,\bar{c}]\; e^{i\int \! d^2x {\cal
L}_{\Delta}}\cr
{\cal L}_{\Delta} &=& \bar{\chi}i \biggl( \gamma^+ P_+ (\partial_+ -i
U^\dagger (L_+ + \Delta b) U + U^\dagger \partial_+ U ) +
\gamma^- P_- D^R_- \biggr) \chi \cr
&& +\frac{1}{4\pi} tr\; \partial_+ U \partial_- U^\dagger +
\frac{1}{4\pi} \int_0^1 \! ds\; tr\; 2i\theta [U^\dagger_s \partial_+
U_s, U^\dagger_s \partial_- U_s ]\cr
&& -\frac{i}{2\pi} tr\; (L_+ - (1-\Delta) b) U D^R_- U^\dagger
-\frac{i}{2\pi} tr\; R_- U^\dagger \partial_+ U \cr
&& +\frac{1}{2\pi} tr\; U^\dagger \bar{c} U D_-^R c ~.
\label{eq:Zgf}
\eea
This functional integral is the desired representation of the theory, which
interpolates
between a purely fermionic and a purely bosonic formulation. We have derived
it here using a slight variant of the Faddeev-Popov procedure, and since it
involves a number of unusual ingredients it is worthwhile to confirm the
derivation from a different point of view.

The idea is to use eq. (\ref{eq:shift}) as the
guide for gauge fixing. We wish to find a gauge such that the purely bosonic
part of (\ref{eq:shift}), $i.e.$
$$
-\frac{i}{2\pi}tr[T_AUD^R_-U^\dagger] ~,
$$
carries a fraction  $\Delta$ of the full
physical current $\bar{\psi}\gamma^+P_+T_A\psi$. In the case of $\Delta\!=\!
1$ this implies that the whole fermionic part of the rotated current
vanishes entirely, i.e. that we are effectively inserting a $\delta$-function
constraint setting this object to zero. Since this object, classically, is
completely gauge invariant, an unusual phenomenon is clearly taking place
here.
The whole gauge-fixing procedure is, in this case ($\Delta\!=\!1$), saved
entirely by quantum effects: quantum mechanically $\bar{\chi}\gamma^+P_+
U^{\dagger}T_AU\chi$ is {\em not} invariant under the gauge transformations
(\ref{eq:sym}). How can we in practice implement such a gauge?
Consider the BRST
procedure. Using the same ghosts, antighosts, and auxiliary fields as
introduced in the Faddeev-Popov procedure discussed above, we first write
down the corresponding BRST transformations:
\bea
\delta\chi(x) &~=~& i c(x) P_+ \chi (x)\cr
\delta\bar{\chi} (x) &~=~& -i \bar{\chi} (x) P_- c(x) \cr
\delta U(x) &~=~& -iU(x) c(x)\cr
\delta U^\dagger(x) &~=~& ic(x) U^\dagger (x)\cr
\delta c(x) &~=~& 0\cr
\delta\bar{c}(x) &~=~& b(x) \cr
\delta b(x) &~=~& 0 ~.
\eea

Next, we add to the un-fixed action a term of the form
\beq
\delta\left[\frac{i}{2\pi}tr(\bar{c}UD^R_-U^{\dagger})\right] = \frac{i}{2\pi}
tr bUD^R_-U^{\dagger}  + \frac{1}{2\pi}tr \bar{c}U(D^R_-c)U^{\dagger} ~,
\eeq
which is BRST-exact. This simply fixes, in a standard manner, the bosonic
objects
$$
tr T_AU D^R_-U^{\dagger}
$$
to zero. The action is now no longer locally gauge invariant, but only
invariant under the global BRST transformations.
However, we have not yet reached the gauge we are looking for. To remedy this,
we shift everywhere the sources $L_+$ by an amount $\Delta b$,
$\Delta$
being the real parameter discussed above. In BRST
formalism, such a procedure is certainly legal. First, what we have done
by shifting the sources $L_+$ in this manner preserves BRST symmetry. This
is obvious, since the action is BRST invariant for arbitrary sources $L_+$,
and $b(x)$ is itself inert under BRST transformations. Second,
and equally important,
this procedure does not change the regularization of the path
integral. In general,
the addition of terms involving couplings between
the auxiliary field and the fermions,
even when done in an apparently BRST-invariant manner, will lead to a
mismatch between the functional measure and the action, and BRST invariance
will be spoiled by terms of order $\hbar$. By shifting the source $L_+$
{\em everywhere}, BRST symmetry is guaranteed even at the quantum
level.\footnote{In the present case, where, in the infinite cut-off limit,
the full action is at most linear in the sources $L_+$, the BRST derivation
presented above may seem unnecessary, since the more simple Faddeev-Popov
procedure is adequate. However, when keeping terms of order $1/\Lambda^2$
in the expansion of the Jacobian (where $\Lambda$ is the overall ultraviolet
cut-off), this more general BRST gauge-fixing procedure is by far
superior. It is also the easiest path to derive the corresponding gauge-fixed
action in higher dimensions.}
So this procedure provides a correct BRST gauge fixing of the path integral,
and it is seen to coincide with the expression (\ref{eq:Zgf}) above.
The rest of this section is concerned with showing that it also provides
precisely the smooth gauge that interpolates between fermions and bosons.

As the argument is presented above,
it is far from obvious that what we have really reached is
a gauge that fixes a certain fraction $\Delta$ of the physical
non-Abelian current $\bar{\psi}\gamma^+P_+T_A\psi$ to be given by the
bosonic current $(i/2\pi)tr T_AUD^R_-U^{\dagger}$. To see that it is,
we carry out explicitly the integration over the auxiliary field $b(x)$.
This leads to  $\delta$-functions inside the functional integral, implementing
the constraints
\beq
\Delta\bar{\chi}\gamma^+P_+U^{\dagger}T_AU\chi + (1-\Delta)\frac{i}{2\pi}
tr T_A UD^R_-U^{\dagger} = 0 ~,
\eeq
which, when comparing with eq. (\ref{eq:shift}), is just what is
required to ensure that the bosonic current is a fraction $\Delta$ of
the full physical current.

The gauge above thus smoothly bosonizes the current $\bar{\psi}\gamma^+P_+
T_A\psi$, but this is of course not the same as bosonizing the full theory.
What about the currents coupled to $R_-$, and what
about the coupling of $U$ to the ghost fields?

A number of miracles conspire to also:

\vspace{0.3cm}

\begin{itemize}

\item Decouple, after a field redefinition, the ghost term.

\item Eliminate the coupling to the sources $R_-$ in the ``boson gauge''
$\Delta\!=\!1$.

\end{itemize}

\vspace{0.3cm}

The key  is vector current
conservation. Let us first write the external sources as
\beq
L_+ = i l^\dagger \partial_+ l~~,~~R_- = i r^\dagger \partial_- r~~.
\label{eq:deflr}
\eeq
Unless $L_+$ and $R_-$ have some global topological properties\footnote{For
example, in the $N$-flavour Schwinger model part of the Abelian
component of the
external vector source $V_\mu$ can be identified with the gauge
field. As already noted in \cite{us}, this field may carry topological
charge, and hence cannot be expressed simply as $V_\mu = \epsilon_{\mu\nu}
\partial^\nu \sigma + \partial_\mu \phi$. Instead a constant component
can be added \cite{us}.}, such rewritings are always possible.

Now we perform  a set of unitary transformations, namely
\bea
\bar{c} &\to& U r^\dagger \bar{c} r U^\dagger \cr
c &\to& r^\dagger c r
\eea
which turn the Lagrangian ${\cal L}_\Delta$ in (\ref{eq:Zgf}) into
\bea
{\cal L}_{\Delta} &\to& \bar{\chi}i \biggl( \gamma^+ P_+ (\partial_+ -i
\Delta U^\dagger b U + U^\dagger D^L_+ U ) +
\gamma^- P_- D_-^R \biggr) \chi \cr
&& +\frac{1}{4\pi} tr\; \partial_+ U \partial_- U^\dagger +
\frac{1}{4\pi} \int_0^1 \! ds\; tr\; 2i\theta [U^\dagger_s \partial_+
U_s, U^\dagger_s \partial_- U_s ]\cr
&& -\frac{i}{2\pi} tr\; (L_+ - (1-\Delta) b) U
D^R_- U^\dagger -\frac{i}{2\pi} tr\; R_- U^\dagger \partial_+ U \cr
&& \frac{1}{2\pi} tr\; \bar{c} \partial_-  c  ~~.
\label{eq:LDelta}
\eea
The ghosts have decoupled now.

The Lagrangian
(\ref{eq:LDelta}) may appear quite complicated, but in certain gauges
this is deceptive. We will demonstrate below that in the two particular
cases $\Delta\!=\!0$ and $\Delta\!=\!1$, one finds the required
simplifications that were advertised above. We have already seen that
in these two particular cases, the current $\bar{\psi}\gamma^+P_+T_A\psi$
is represented either entirely in terms of a fermionic current, or
a bosonic current. We thus expect that in the gauge $\Delta\!=\!0$
we are dealing with a completely fermionic representation, and in the
gauge $\Delta\!=\!1$ with a completely bosonic representation of the
same theory. This is far from obvious, in both cases, since the
gauge-fixing functions in this non-Abelian case remove entirely neither the
fermionic fields, nor the bosonic fields.

Since all fields are always present in the path integral, the proof of
fermionization ($\Delta\!=\!0$) and bosonization ($\Delta\!=\!1$) will
rely on showing that in these two limiting cases we can decouple
either the bosonic or the fermionic fields from the external sources
$L_+$ and $R_-$. The easiest way is to go back to
the starting expression in terms of the action (\ref{eq:Spsi}).
Note that a pure non-Abelian phase rotation,
$$\psi \to r^\dagger \psi~~, \bar{\psi} \to \bar{\psi}r~~,$$
eliminates the coupling to $R_-$ at the expense of providing a
modified source $\bar{L}_+$:
\beq
L_+ \to \bar{L}_+ \equiv irl^\dagger \partial_+ (lr^\dagger)~~.
\eeq
This means that as long as (\ref{eq:deflr}) is a valid substitution
for the external sources, we can equally well work with just a source
$\bar{L}_+$ defined as above (replacing $L_+$ in eq. (\ref{eq:Spsi})),
and then  set $R_-$ equal to zero.

With that in mind, let us first concentrate on the case $\Delta =0$.
In this case the auxiliary field $b(x)$ only couples to the bosonic
field $U(x)$. But the source $\bar{L}_+$ still couples to the bosonic
fields as well, so it may appear as if this is not yet the sought-for
fermionic representation of the theory. Could this be true? We
have introduced the collective fields $U(x)$ redundantly in the path
integral, and have, in this particular gauge, put one simple
gauge-fixing condition on these new fields. Clearly such a procedure
cannot introduce physical couplings to the redundant fields $U(x)$.
This means that the couplings between $\bar{L}_+$ and $U(x)$ must be
illusory, cancelling out whenever we compute physical quantities.

To see that this indeed is the case, we first perform a phase rotation
of the fermion fields:
\beq
\chi \to U^\dagger \chi ~~,~~\bar{\chi} \to \bar{\chi} U~~.
\eeq
The resulting couplings of $U(x)$ are then
\beq
\frac{i}{2\pi} tr\; (b - \bar{L}_+ ) U \partial_- U^\dagger +
i\bar{\chi} \gamma^- P_- U\partial_- U^\dagger \chi~~.
\eeq
At this point it is important that, at least for the Lie groups we are
considering, quantities like $U \partial_- U^\dagger $ can be
expressed as a linear combination of the generators $T_A$ of the Lie
group. Then, obviously, a shift
$$ b^A \to b^A + \bar{L}_+^A - \frac{2\pi}{\alpha} \bar{\chi} \gamma^- P_-
T^A \chi~, $$
where $\alpha\delta_{AB}\!=\!trT_AT_B$, eliminates the couplings
of $U(x)$ to both the external source $\bar{L}_+$, and to the
fermions. The constant $\alpha$ is here defined through the relation $tr
T_A T_B  = \alpha \delta_{AB} $. The resulting Lagrangian, after a
transformation $\chi \to r \chi~,\bar{\chi} \to \bar{\chi} r^\dagger$
is, on account of (\ref{eq:deflr}):
\bea
{\cal L}_{\Delta =0} &\to& \bar{\chi}i \biggl( \gamma^+ P_+ D^L_+  +
\gamma^- P_- D^R_- \biggr) \chi \cr
&& +\frac{1}{4\pi} tr\; \partial_+ U \partial_- U^\dagger +
\frac{1}{4\pi} \int_0^1 \! ds\; tr\; 2i\theta [U^\dagger_s \partial_+
U_s, U^\dagger_s \partial_- U_s ]\cr
&& +\frac{i}{2\pi} tr\; b U \partial_- U^\dagger + \frac{1}{2\pi} tr\;
\bar{c} \partial_-  c  ~~.
\label{eq:Lf}
\eea
This shows explicitly that
the Green functions in this gauge are given entirely by the fermionic
functional integral. The integration over $b, U$ and the ghosts
$\bar{c},c$ simply yields an unimportant normalization factor.

For $\Delta\!=\!1$, the auxiliary field $b(x)$ only couples to the
fermions. We may now shift $b$ by $i U D^{\bar{L}}_+ U^\dagger $,
thereby eliminating any coupling of the fermions to $U$ or the
external sources. Here, again, we make use of the fact that $i U
D^{\bar{L}}_+ U^\dagger $ may be expressed as a linear combination of
the generators $T_A$. In order to recover the original sources $L_+$ and
$R_-$ again, we transform $U \to r U r^\dagger$ and use the definitions
(\ref{eq:deflr}).
The functional integration over $\chi$,$\bar{\chi}$,$b$,$c$
and $\bar{c}$ will then only yield a normalization factor ${\cal N}$ and
the final result is a generating functional
\bea
Z[V,A] &=& {\cal N} \int\! {\cal D}[U]\; e^{i\int \!
d^2x {\cal L}_{bos}}\cr
{\cal L}_{bos} &=& +\frac{1}{4\pi} tr\; \partial_+ U \partial_- U^\dagger +
\frac{1}{4\pi} \int_0^1 \! ds\; tr\; 2i\theta [U^\dagger_s \partial_+
U_s, U^\dagger_s \partial_- U_s ]\cr
&& -\frac{i}{2\pi} tr\; L_+  U D^R_- U^\dagger
-\frac{i}{2\pi} tr\; R_- U^\dagger \partial_+ U ~~.
\label{eq:Zbos}
\eea
This constitutes the desired bosonic representation of the original
functional integral (\ref{eq:Zf}). Note that, just as in the Abelian case,
the fermions have not been gauged completely away in this formulation.
They only happen to
decouple from the external sources in this gauge, having neither
(non-Abelian) vector nor axial vector currents. For this reason it is
convenient to simply integrate them out, together with the ghosts and
auxiliary fields.

By taking functional derivatives with respect to the external sources,
we read off the effective bosonization rules from the representation
(\ref{eq:Zbos}):
\bea
\bar{\psi}\gamma^+P_+T_A\psi &~\sim~& -\frac{i}{2\pi}tr T_AU\partial_-
U^{\dagger} \cr
\bar{\psi}\gamma^-P_-T_A\psi &~\sim~& -\frac{i}{2\pi}tr T_AU^\dagger
\partial_+U ~.
\eea
These bosonization rules, and the bosonized action (\ref{eq:Zbos}),
coincide with
those derived by Witten \cite{Witten}. When products of currents are
taken, it follows from (\ref{eq:Zbos}) that additional ``contact terms'',
originating from the part involving $L_+$ and $R_-$ simultaneously, may
arise. Also these terms, which depend on the regularization chosen,
are well understood \cite{DiVecchia}. If one considers commutators of
currents, the contact terms do not contribute, and the above simple current
bosonization rules suffice.

For any other gauge with $\Delta\!\neq\!1$, it is not possible to
write down analogous bosonization relations. The theory is then only
in its partially bosonized form, and there are both fermionic and bosonic
contributions to physical Green functions. As in the Abelian case
\cite{us}, one may even contemplate gauges for which $\Delta$ becomes
a function of the space-time position as well. The existence of
mixed representations which involve both fermions and bosons in a
non-trivial interactive manner could not be inferred from the
original bosonization arguments.

As in the Abelian case, it is straightforward to introduce genuine
interaction terms in fermionic language, and use the method of smooth
bosonization to find their bosonized or partly bosonized equivalents.
This is done, for vector and axial vector interactions, by treating
(part of) the sources $L_+$ and $R_-$ as dynamical degrees of freedom,
to be integrated over in the functional integral with an appropriate
measure. In this manner one can immediately treat two-dimensional
thories such as the non-Abelian Thirring model and
$N_f$-flavour QCD$_2$ (of an arbitrary number of colours $N_c$).

\section{Comments and conclusions}

Our first comment concerns the regularization
of the original fermionic path integral,
and the terms of order ${\cal O} (\Lambda^{-2})$ in the effective
action. The structure of these additional
terms is similar to the
Abelian case, and in both (1+1) and (3+1) dimensions \cite{us,us1}.
The higher
derivatives acting on $U(x)$ can thus be interpreted as an induced
``regularization'' on these bosonic fields. Such a regularization is
to be expected, because
we started with a regularized functional integral and have performed
manipulations which should not change this feature. The terms involving
higher powers of the external sources are certainly suppressed in the
limit $\Lambda \to \infty$, except perhaps for pathological cases where these
sources are taken to be of the order of the cutoff. Even if these
external sources are taken to be dynamical fields, this will not
happen; any gauge theory in (1+1) dimensions has an explicit scale
set by the coupling constant $g$, and not by the ultraviolet cutoff
$\Lambda$. It should be noted, however, that removing these extra
terms in the formal limit $\Lambda \to \infty$ is not entirely as
simple as it may appear on the surface, even here in (1+1)
dimensions. In particular, if
mass terms are included the cut-off--dependent terms turn out to play
a crucial r\^{o}le. See ref. \cite{us} for a discussion of the
corresponding Abelian case.

Another comment concerns the choice of gauge as implemented in
(\ref{eq:Zgf}) which smoothly
interpolates between fermionic (at $\Delta \!=\! 0$) and bosonic (at $\Delta
\!=\! 1$) descriptions. This gauge is well defined up to a zero mode, as is
obvious if we consider the special case of $\Delta\!=\! 0$ in the
formulation in which the source $R_-$ has been removed by a phase
rotation (and then effectively included in the modified source
$\bar{L}_+$). Here it clearly
fixes only a combination involving the {\em derivative} of $U$ to zero :
$tr T_A U\partial_-U^\dagger = 0$.\footnote{The zero mode is
in fact present for
all values of $\Delta$. See the analogous discussion for the Abelian case
in ref. \cite{us}.} So a space-time--independent $U$-component remains
unfixed.
For the massless case we have been considering here,
this incompleteness in the chosen gauge fixing is of no concern, but it
is crucial to gauge away this zero mode if one wishes to treat also the
massive case correctly (see the second reference of \cite{us}).

To conclude, we have provided one possible non-Abelian generalization
of smooth bosonization. In the particular case of the bosonization
gauge $\Delta\!=\!1$, we find an entirely bosonic theory based on a
sigma model with a Wess-Zumino term, while in the fermionic gauge
$\Delta\!=\!0$, we find a purely fermionic theory. By tracing the
coupling to external sources $L_+$ and $R_-$, we regain the
non-Abelian bosonization rules of Witten. For all other values of the
gauge parameter $\Delta$ we obtain mixed representations of fermions
and bosons. All of these theories are equivalent.

A natural question at this stage concerns the generalization of this
bosonization prescription to higher dimensions. It should be obvious
that a number of relations, very particular
to two space-time dimensions, conspire to ensure the existence of a
(local) purely bosonic gauge when a purely fermionic gauge exists. It
is not inconceivable that three-dimensional bosonization based on the
introduction of a Chern-Simons term \cite{Luscher} can be shown
similarly to be part of a larger class of equivalence. In four dimensions
the hopes for a local bosonic quantum field theory being entirely
equivalent to a local fermionic theory are small. This does not imply
that various stages of ``partial bosonization'' cannot be achieved. In
fact, the present formalism is ideally suited for a formulation of
theories with fermions (such as QCD) in partly bosonized terms. The
gauge-fixing function then stipulates precisely which objects can be
represented in bosonic variables. Some examples have been given in
refs. \cite{us1,us2}.

\vspace{0.5cm}

\noindent
{\sc Acknowledgement:} ~We both acknowledge encouraging and
stimulating discussions with Holger Bech Nielsen.
One of us (P.H.D.) would also like to thank Cliff
Burgess and Fernando Quevedo for discussions.
We finally wish to thank D.M. Brink for giving us the opportunity
to enjoy a pleasant stay at the European Centre for
Theoretical Studies in Nuclear Physics and Related Areas, ECT*, in
Trento, Italy, where the work described here was completed. This
work was also supported in part by Norfa grant no. 93.15.078/00.

\end{document}